\documentclass{svproc}
\usepackage{changepage}
\usepackage{subcaption}
\usepackage{comment}
\usepackage{graphicx}%
\usepackage{multirow}%
\usepackage{amsmath,amssymb,amsfonts}%
\usepackage{mathrsfs}%
\usepackage[title]{appendix}%
\usepackage{xcolor}%
\usepackage{textcomp}%
\usepackage{manyfoot}%
\usepackage{booktabs}%
\usepackage{algorithm}%
\usepackage{algorithmicx}%
\usepackage{algpseudocode}%
\usepackage{listings}%
\usepackage{url}

\begin{document}
\mainmatter              
\title{Influence of Assortativity on the SICP Spreading Process in Hypergraphs}
\titlerunning{Influence of Hypergraph Assortativity on SICP}  
%
\author{Jelmer Stolwijk\inst{1} \and Renze Dijkhuizen\inst{1} \and H.A. (Bart) Peters\inst{1} \and Huijuan Wang\inst{1}}
\authorrunning{Jelmer Stolwijk et al.} 

\institute{$^1$Faculty of Electrical Engineering, Mathematics, and Computer Science,\\ Delft University of Technology, The Netherlands\\
}

\maketitle              

\begin{abstract}
Many real-world systems involve group interactions that cannot be captured by pairwise graphs. They can be better represented by hypergraphs where each hyperedge could connect an arbitrary number of nodes. The Susceptible–Infected Contact Process (SICP) has been proposed to model the spreading of information and epidemics on hypergraphs. However, how higher-order topological properties of a hypergraph influence the SICP spreading process remains understudied. In this paper, we examine the impact of hypergraph assortativity based on the Neighborhood Coefficient (NC)—the average size of hyperedges incident to a node—on the SICP spreading process. By introducing a low-complexity measure of hypergraph assortativity grounded in analysis of variance, we apply hyperdegree-preserving rewirings to real-world networks and synthetic hypergraph models to tune their assortativity. We find and explain why lower NC assortativity yields higher average epidemic prevalence and lower prevalence variance across different seed selections. This phenomenon is evident when hyperdegree-preserving rewirings are applied to real-world networks with heterogeneous topological structures, but it is largely absent when comparing synthetic random hypergraphs to their rewired counterparts.

\end{abstract}

\section{Introduction}\label{introduction}

Social networks facilitate the spread of information and epidemics. Understanding the influence of the underlying network on a spreading process is crucial for the estimation and mitigation of epidemic/misinformation spreading. Many real-world systems involve group interactions that cannot be precisely represented by classic pairwise networks, where each edge connects two nodes. For example, a face-to-face interaction, online discussion, or scientific collaboration could engage more than two individuals. Hence, hypergraphs, where nodes represent entities and hyperedges represent the relationship of two or more nodes simultaneously, have recently been studied \cite{10.1371/journal.pone.0323753,BATTISTON20201,battiston2021physics,ceria2023temporal,Zhangbackbone}.

A new challenge is understand how topological properties of the underlying hypergraphs influence dynamic processes unfolding on these networks. Recent work has begun to address this challenge by evaluating spreading processes on synthetic hypergraph models with a given topological property \cite{Chodrow2020ConfigModelsRandomHypergraphs,Landry2020EffectheterogeneityHypergraphContagion}. This challenge has also been investigated by evaluating dynamics on real-world networks alongside rewired counterparts that adjust target properties, such as assortativity, while preserving the original hyperdegree sequence $\{d_i^{k}\}$ of each node, where $d_i^{k}$ denotes the number of hyperedges containing $k$ nodes that include $i$ \cite{Landry2022HypergraphAssortativityDynamicalSystems}. While continuous-time hypergraph Susceptible–Infected-Susceptible models have been explored in previous works, discrete-time dynamics under the Susceptible–Infected Contact Process (SICP) have been largely restricted to influence maximization problems (i.e., identifying optimal seed sets for maximum prevalence) \cite{abhelakh2026fairinfluencemaximizationhypergraphs,gong_influence_2024,xie_efficient_2023}. How the SICP spreading process depends on the underlying topological properties of the hypergraph remains significantly less investigated. 

To bridge this gap, we will explore the influence of assortativity on SICP spreading as follows. First, we adopt the hypergraph rewiring method proposed in \cite{Landry2022HypergraphAssortativityDynamicalSystems} that systematically changes a targeting network property of a given hypergraph while preserving the order-specific hyperdegree $d_i^{k}$ 
of each node $i$ at each order $k$ in the given hypergraph. 
Second, the target property to vary by rewirings, namely the assortativity of any nodal property in hypergraphs, will be defined. Assortativity describes the tendency of similar nodes being connected by a hyperlink, where the similarity of nodes may be defined by any property. Assortativity was originally formulated for pairwise networks, traditionally using Pearson correlation coefficients to measure the similarity of linked nodes \cite{newman_mixing_2003}. Extending assortativity to hypergraphs is non-trivial: projecting hyperedges into pairwise cliques quadratically overweights large hyperedges \cite{Landry2022HypergraphAssortativityDynamicalSystems}, while more computationally intensive definitions rely on choice functions \cite{Chodrow2020ConfigModelsRandomHypergraphs}. Here, we introduce an intuitive and low-complexity measure of hypergraph assortativity. Grounded in classical analysis of variance \cite{Fisher1925StatisticalMethods}, our definition offers a clear physical interpretation and enables computationally efficient hyperdegree preserving rewirings to increase or decrease assortativity.
Third, we study the influence of assortativity on SICP by rewiring real-world as well as synthetic hypergraphs to increase or decrease assortativity. 
The SICP spreading process starts at $t=0$ with a single infected/informed node, while the other nodes are susceptible. At each time step, every infected node selects one of the hyperedges it belongs to uniformly at random, and infects each susceptible node in that hyperedge with an infection probability $\beta$. The process terminates after a time step $T$. 

For pairwise graphs, assortativity of two end nodes of an edge in degree may significantly affect spreading processes on pairwise networks \cite{PhysRevE.76.026116,van_mieghem_influence_2010}. For SICP spreading on hypergraphs, a node with a high Neighborhood Coefficient, defined as the average size of the hyperedges incident to the node, tends to result in higher prevalence when selected as the single seed node \cite{gong_influence_2024}. 
The intuition behind this finding is that as every infected node randomly selects one of its incident hyperedges and tries to infect each susceptible node within that hyperedge, a high Neighborhood Coefficient implies more susceptible nodes in the chosen hyperedge to infect. Hence, we will apply hyperdegree preserving to real-world hypergraphs as well as hypergraphs generated by models to tune the assortativity by Neighborhood Coefficient and explore the influence of this assortativity on the SICP spreading process.

In hypergraphs with heterogeneous structure such as most real-world networks, we find that lower assortativity tends to lead to a higher prevalence on average and lower variation in prevalence introduced by different choices of the single seed node. Our method could be applied broadly to understand the influence of assortativity on generic dynamics processes on hypergraphs.

\section{Hyperdegree sequence preserving rewiring}
A hypergraph can be defined as $H=(V,E)$, where $V$ denotes the set of $|V|$ nodes and $E$ the set of $|E|$ hyperedges, respectively. A hyperedge $e_i \in E$ can connect any number of nodes greater than or equal to $2$. Given a hypergraph $H$, a hyperdegree preserving rewiring modifies the given hypergraph while the hyperedge degree $d_i^{k}$ of each node $i$ at each order $k$ is preserved. Here, the hyperedge degree $d_i^{k}$ describes the number of hyperedges of order $k$ that include $i$, where the order $k$ of a hyperedge refers to the number of nodes in the hyperedge. 

At each rewiring step, two hyperedges $e_1,e_2 \in E$ of equal order are selected uniformly at random. A node is then selected uniformly at random from each hyperedge, i.e., node $u \in e_1$ and node $v \in e_2$. These two nodes are swapped:
\begin{equation*}
    e_1 \leftarrow (e_1 \setminus \{u\}) \cup \{v\}, \qquad
    e_2 \leftarrow (e_2 \setminus \{v\}) \cup \{u\},
\end{equation*}
provided that the resulting hyperedges preserve their original size. If the swap fails the condition, the selection of hyperedges and their respective nodes is repeated until a valid swap is found. Duplicate hyperedges are allowed to exist in the hypergraph. This node swap rewires strictly two hyperedges, preserving the exact node composition of all remaining hyperedges in the hypergraph. This rewiring step without any control of network property can be repeatedly applied to any given hypergraph, randomizing the connections. By selecting two hyperedges of equal order and performing node swapping, the hyperdegree of every node is preserved across all orders.

This rewiring step can also be generalized to control a network property, assortativity in our case. This generalization is achieved by adding an additional condition when swapping nodes: the two nodes $u$ and $v$ are swapped, provided that the resulting hyperedges preserve their original size and the resulting hypergraph has a higher (lower) assortativity, if the objective is to increase (decrease) assortativity.

\section{Assortativity} \label{strategies}
We focus on any nodal property $f(v) \in \mathbb{R}$, whose global multiset of node attributes $\mathcal{F} = \{\{ f(v) \mid v \in V \}\}$ remains strictly invariant under the proposed rewiring process. Examples include non-topological traits like individual age, and topological properties such as the neighborhood coefficient.
Our definition is based on a traditional analysis of variance. First, the within-hyperedge variation of property $f$ across all hyperedges is defined as 
$\sum_{e \in E}\sum_{v\in e}\left(f(v)-\overline{f}(e)\right)^2$, where $\overline{f}(e)$ denotes the average of property $f$ over all nodes in hyperedge $e$. 
We then define assortativity $r$ with respect to a nodal property $f$ as
\begin{equation}
\label{eq:r}
    r=\frac{\sum_{e \in E}\sum_{v \in e}{(f (v) - \mu)^2}-\sum_{e \in E}\sum_{v\in e}\left(f(v)-\overline{f}(e)\right)^2}{\sum_{e \in E}\sum_{v \in e}{(f (v) - \mu)^2}},
\end{equation}
where $\mu = \frac{\sum_{e \in E}\sum_{v \in e}f(v)}{\sum_{e \in E}\sum_{v \in e}1}$ represents the average property $f$ across all node-hyperedge incidences. Equivalently, $\mu =\frac{\sum_{v\in V}f(v)\cdot D_v}{\sum_{v\in V}D_v}$ is the average property $f$ across all nodes where each node is weighted by its hyperedge degree $D_v=\sum_{k=2}^{\infty} d_v^k$, i.e., its total number of hyperedges.

Assortativity $r$ then represents the fraction of total variance explained by hyperedge grouping. When $\sum_{v\in e}\left(f(v)-\overline{f}(e)\right)^2\to 0$, i.e., nodes in every hyperedge $e$ are almost identical in property $f$, and  $r \to 1$ the hypergraph exhibits maximum assortativity/homophily. Conversely, $r=0$ means that nodes inside a hyperedge vary as much as the entire global network, indicating random mixing. Naturally, $r\in[0,1]$, because
\begin{equation}
\begin{aligned}
\sum_{e \in E}\sum_{v \in e}(f(v) - \mu)^2 
&= \sum_{e \in E}\sum_{v \in e}\left[\left(f(v) - \overline{f}(e)\right) + \left(\overline{f}(e) - \mu\right)\right]^2 \\
&= \sum_{e \in E}\sum_{v \in e}\left(f(v) - \overline{f}(e)\right)^2 + \sum_{e \in E}|e|\left(\overline{f}(e) - \mu\right)^2 \\
&\geq \sum_{e \in E}\sum_{v \in e}\left(f(v) - \overline{f}(e)\right)^2.
\end{aligned}
\end{equation}

When two nodes are swapped during the hyperdegree sequence preserving rewiring, only the term $\sum_{v\in e}\left(f(v)-\overline{f}(e)\right)^2$ for the two affected hyperedges could change. The other term in eq. \ref{eq:r} is constant, because 
\[
\sum_{e \in E}\sum_{v \in e}{(f (v) - \mu)^2}=\sum_{v\in V}D_v(f(v)-\mu)^2,
\]
and our choice of nodal properties is such that the multiset of node attributes remains invariant under the rewiring process.

\section{Effect of Rewiring on SICP spreading process} \label{res}
We apply the proposed rewiring methods to understand the influence of assortativity by neighborhood coefficient on the SICP. The proposed rewiring method is applied to real-world networks to increase and decrease assortativity. Performance of the SICP on these networks including rewired ones are evaluated. 

In our analysis, we will consider several real-world hypergraphs, of which the topological properties are shown in Table \ref{table:datasets}. In the Algebra \& Geometry hypergraphs \cite{alg_geo_res_bar_amburg2020fair}, hyperedges correspond to groups of MathOverflow users responding to the same question related to algebra and geometry, respectively. The nodes in the Restaurants-Rev and Bars-Rev hypergraphs \cite{alg_geo_res_bar_amburg2020fair} represent Yelp users, who can form a hyperedge if they have reviewed the same restaurant or bar category, respectively, within a month. Similarly, the hyperedges in the Music-Rev hypergraph \cite{mus_ni2019justifying} represent groups of Amazon users that have reviewed the same blues-music category in the same month. Lastly, the hyperedges in the contact-primary-school hypergraph \cite{pri1_Benson-2018-simplicial,pri2_Stehl-2011-contact} correspond to groups of students in French schools, where the students in each group have simultaneously been in close proximity of each other at the same time.
\begin{table}[t]
    \centering
    \captionsetup{labelfont=bf}
    \caption{Statistics of the datasets used in this study. $|V|$ denotes the number of nodes, $|E|$ the number of hyperedges, $|E_3|$ the number of size-3 hyperedges, $|E_4|$ the number of size-4 hyperedges, $|e|_{\max}$ the maximum hyperedge size, $|e|_{\mathrm{\mathrm{avg}}}$ the average hyperedge size, and $E[D]$ the average hyperdegree}
    
    \begin{tabular}{lrrrrrrr}
    Dataset & $|V|$ & $|E|$ & $|E_3|$ & $|E_4|$ & $|e|_{\max}$ & $|e|_{\mathrm{avg}}$ & $E[D]$ \\
    \hline
    Algebra & 423 & 1268 & 205 & 114 & 107 & 6.52 & 19.53 \\
    Geometry & 580 & 1193 & 196 & 131 & 230 & 10.47 & 21.53 \\
    Music-Rev & 1106 & 694 & 52 & 39 & 83 & 15.13 & 9.49 \\
    Restaurants-Rev & 565 & 601 & 95 & 46 & 43 & 7.66 & 8.14 \\
    Bars-Rev & 1234 & 1194 & 191 & 153 & 73 & 9.94 & 9.62 \\
    contact-primary-school & 242 & 12704 & 4600 & 347 & 5 & 2.42 & 126.98 \\
    \end{tabular}
   
    \label{table:datasets}
\end{table}

We will rewire order-3 hyperlinks as an example, and explore the corresponding effect on the SICP spreading process to demonstrate the method. This is also motivated by the larger number of order-3 hyperlinks than higher-order ones, e.g., order-4, as shown in Table \ref{table:datasets}. 

Given a hypergraph, $m$ steps of rewiring are applied to increase and decrease assortativity respectively to obtain a rewired synthetic network, where $m \in \{10\% |E_3|, 20\% |E_3|, 50\% |E_3|\}$ and $|E_3|$ is the number order-3 hyperedges. For a given hypergraph and $m$, 10 synthetic rewired networks are generated independently. 
For the SICP, we consider the infection probability $\beta=0.02$ and the process ends after $T=25$ time steps, a typical choice when the SICP and aforementioned real-world networks are used to study influence maximization processes \cite{abhelakh2026fairinfluencemaximizationhypergraphs,gong_influence_2024,xie_efficient_2023}. 

For a given hypergraph — whether an empirical network or a synthetic variant obtained through $m$ steps of assortative (or disassortative) rewiring — we compute the average number of infected nodes at $T$ for each possible seed node, averaged across $1000$ independent realizations of the SICP spreading process. The prevalence $\sigma$ of the SICP on a hypergraph is the average number of infected nodes, averaged further across all possible single seed nodes. The effect of assortativity on the SICP process is evaluated via the prevalence and the variation of the average number of infected nodes introduced by each seed node.

\subsection{Prevalence}
\begin{figure*}[t]
    \centering

    \includegraphics[width=0.49\textwidth]{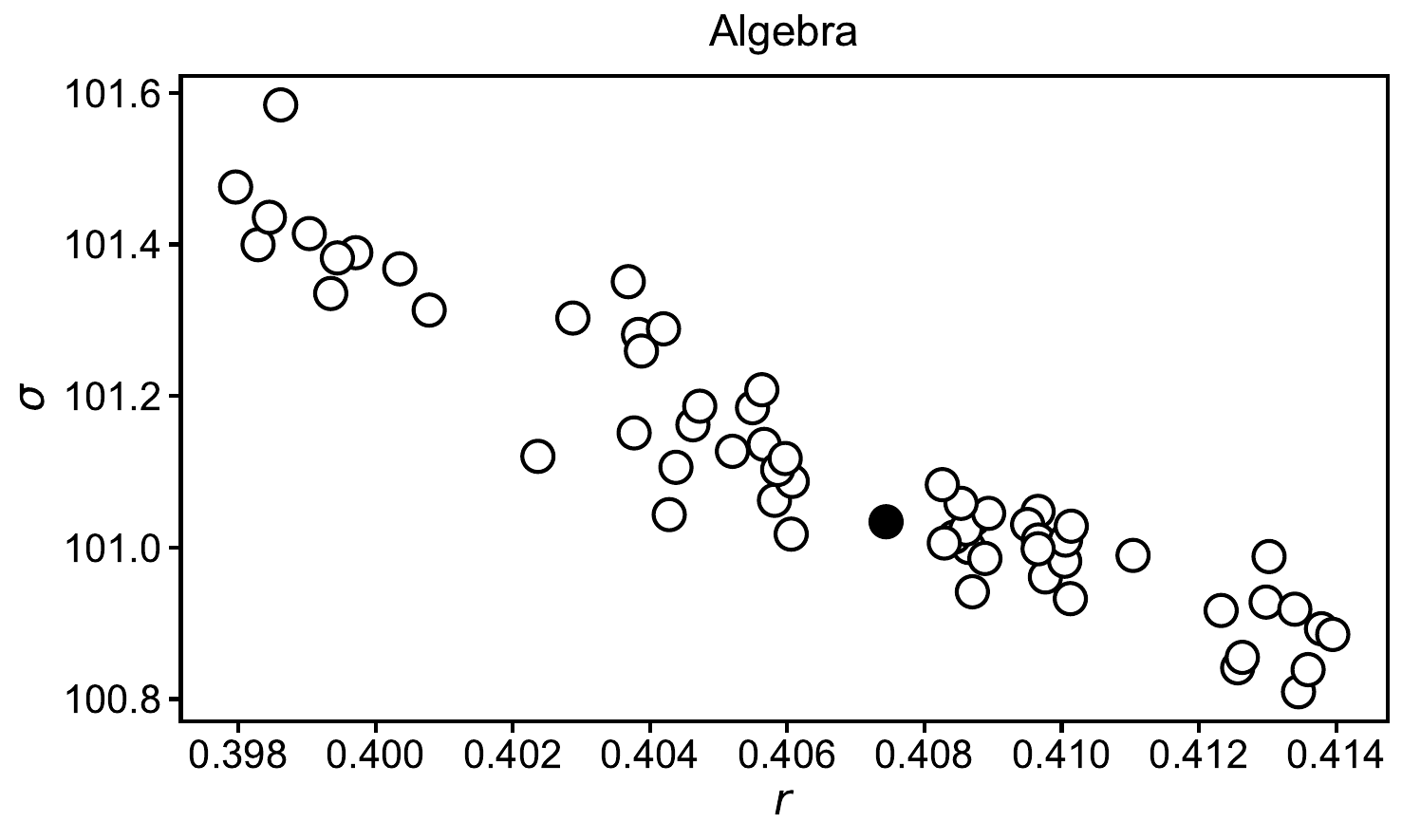}
    \includegraphics[width=0.49\textwidth]{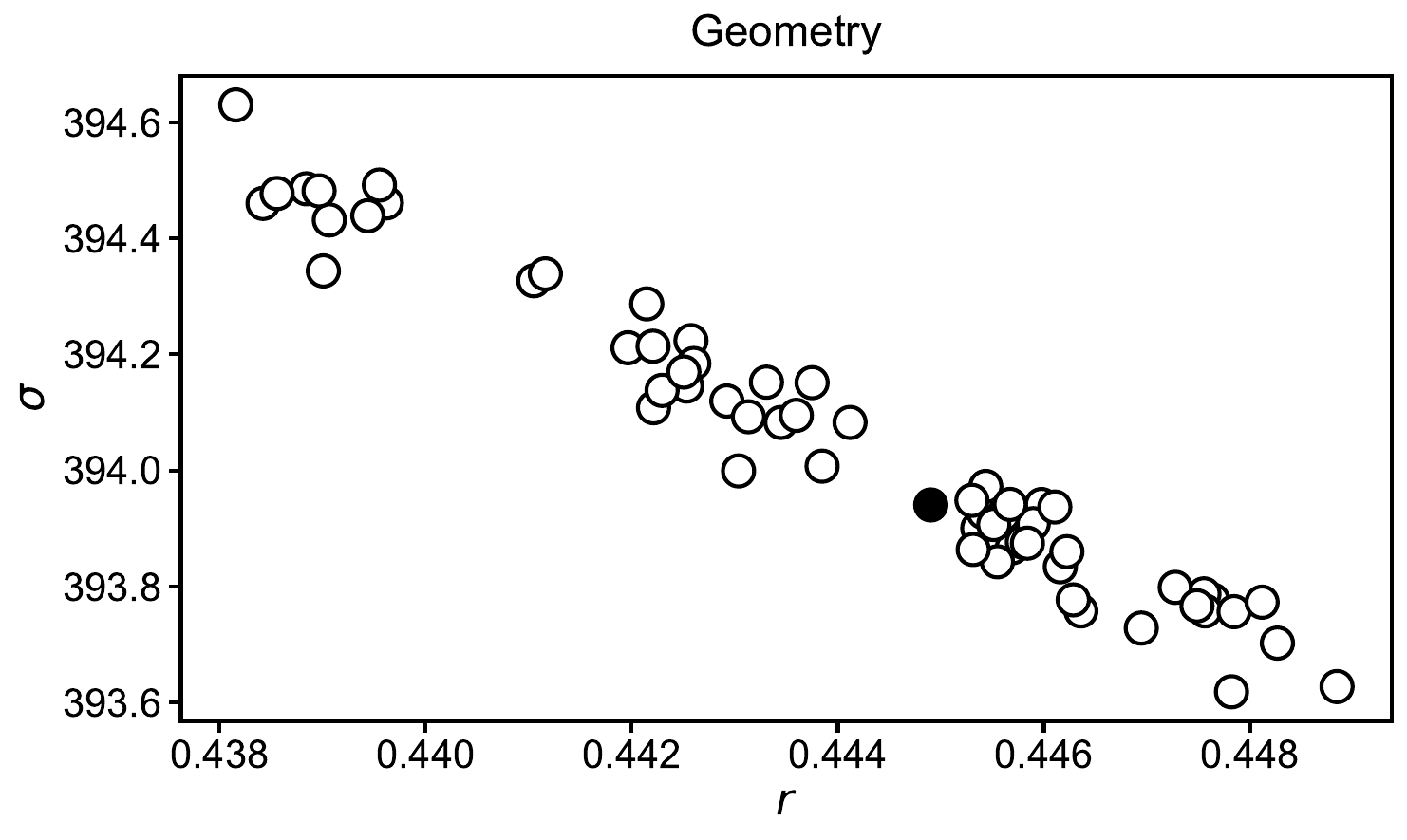}
    \vspace{0.5em}
    \includegraphics[width=0.49\textwidth]{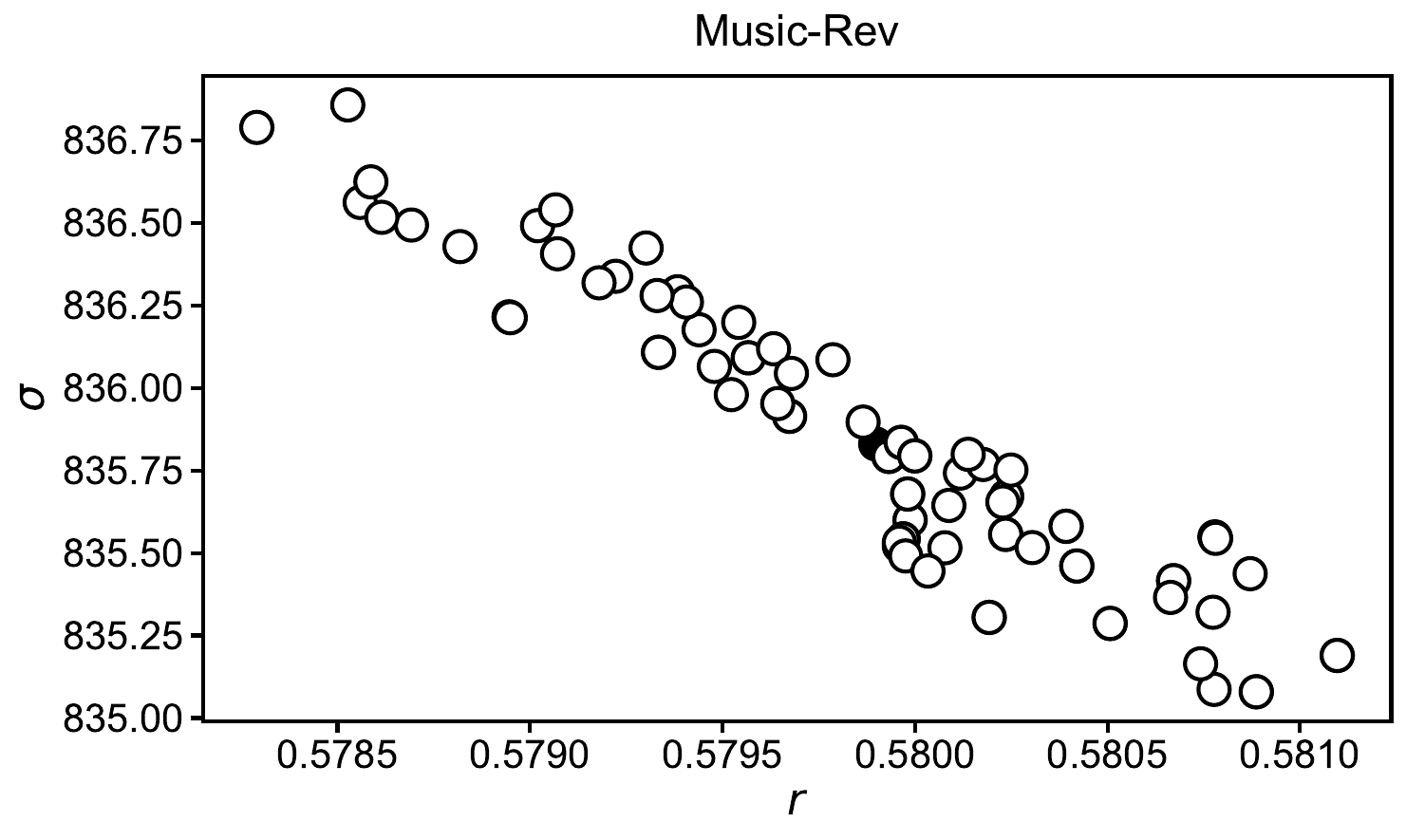}
    \includegraphics[width=0.49\textwidth]{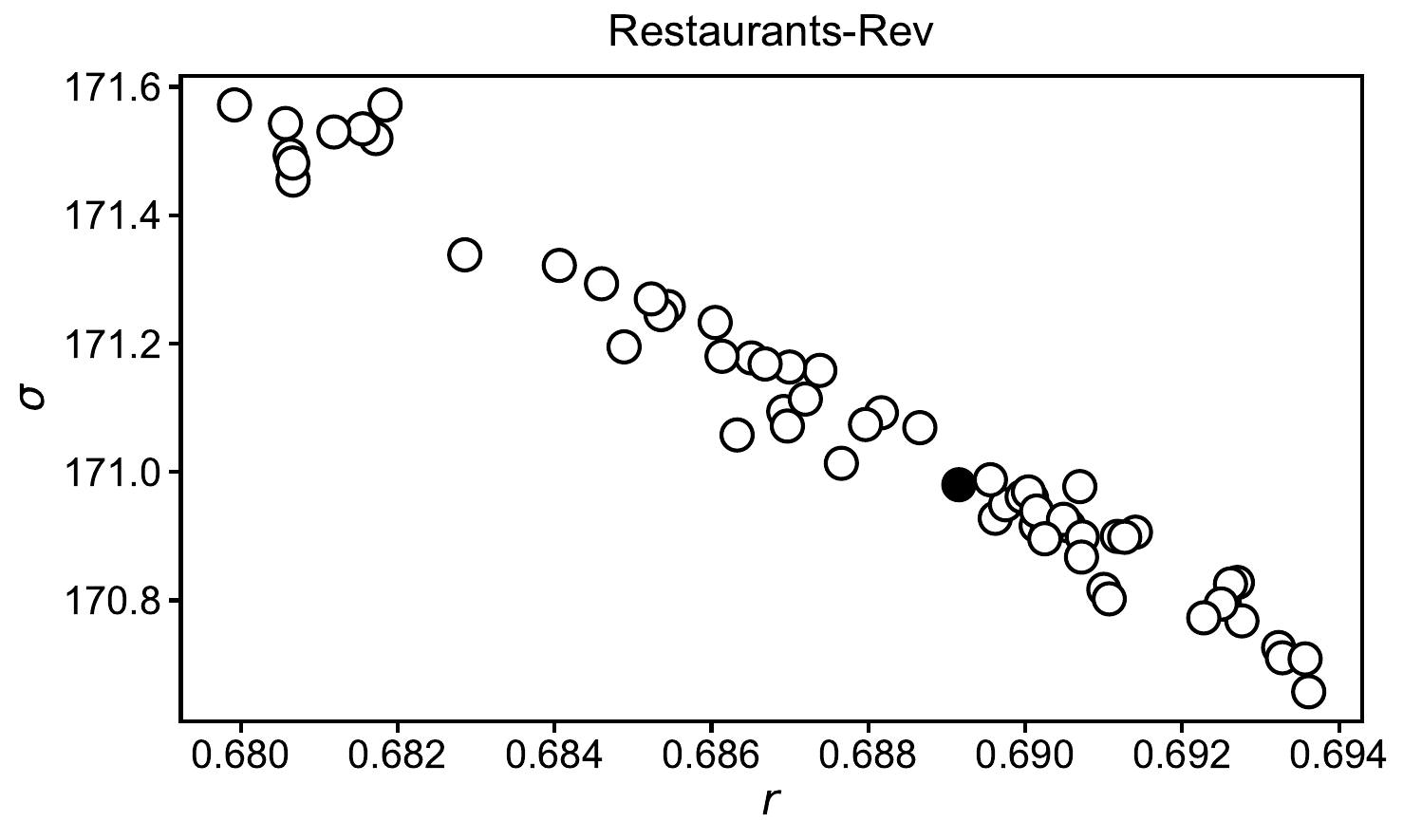}
    \vspace{0.5em}
    \includegraphics[width=0.49\textwidth]{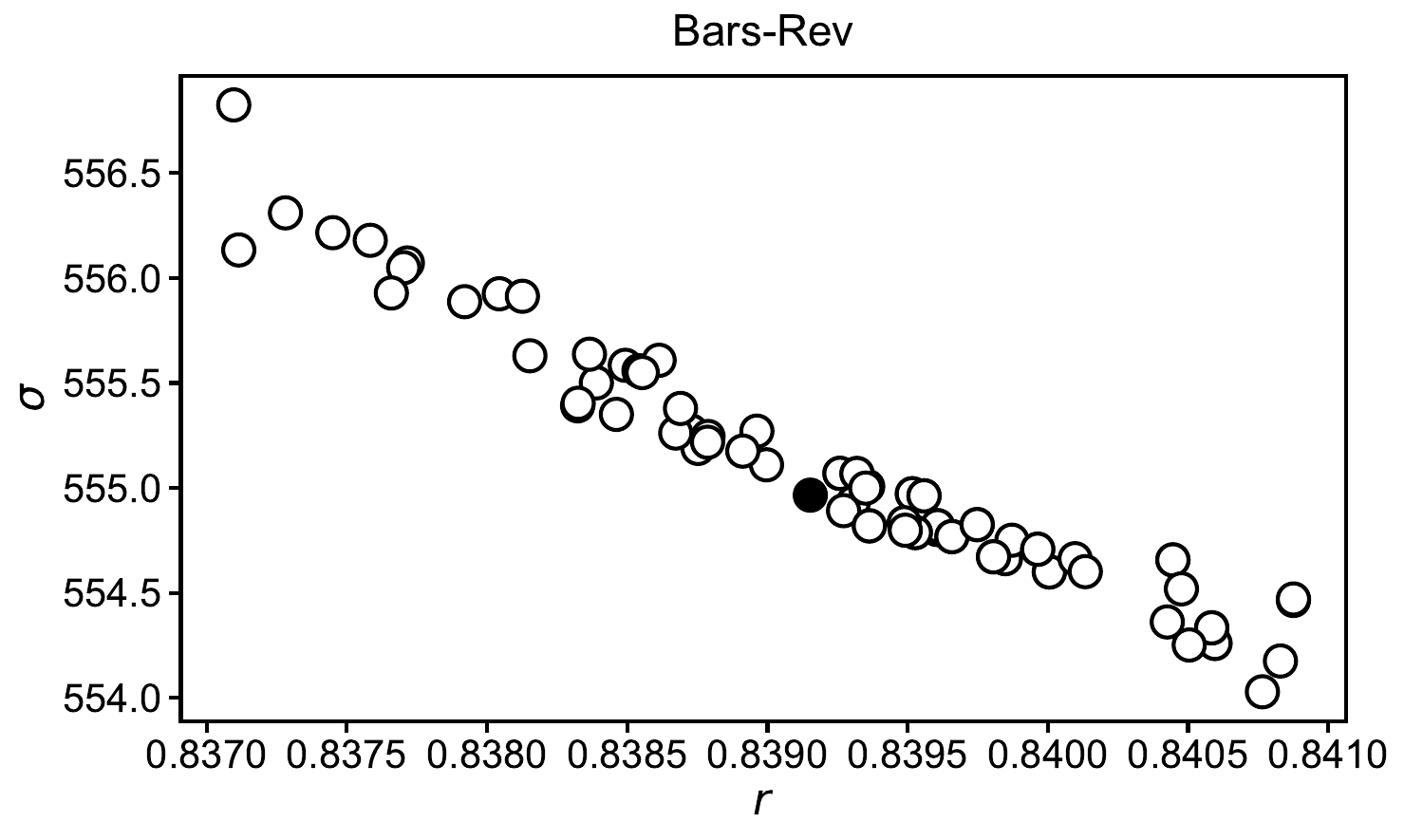}
    \includegraphics[width=0.49\textwidth]{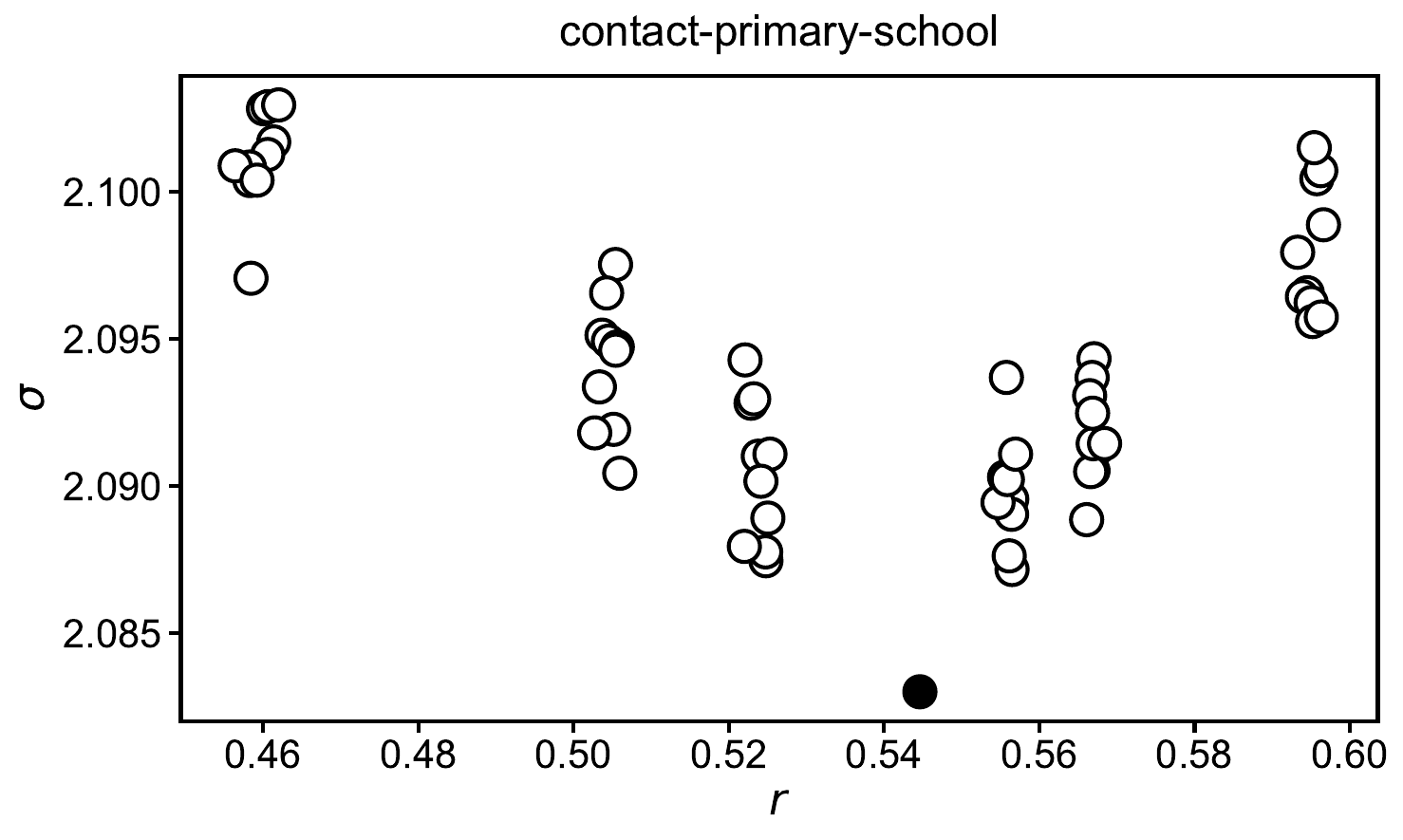}
    
    \captionsetup{labelfont=bf}
    \caption{
        Prevalence $\sigma$ after $T=25$ time steps in relation to NC assortativity $r$ under order-3 hyperedge rewiring. Results are averaged over all seed nodes and simulation runs. The solid black dot denotes the unmodified hypergraph.
    }
    \label{fig:all_datasets_means}
\end{figure*}

Figure~\ref{fig:all_datasets_means} shows the prevalence $\sigma$ of the spreading process versus the assortativity $r$ of a real-world hypergraph and its rewired counterparts. As the hypergraph becomes more disassortative, the prevalence tends to increase. This is observed in all real-world networks except for contact-primary-school. This negative correlation between prevalence and assortativity can be partially explained by the following. 

When any node belonging to a large hyperlink is infected, the hyperlink is possibly selected to infect its remaining susceptible nodes. Hence, a node with many large hyperlinks is more prone to infection. A node with a high neighborhood coefficient (NC, defined as the average size of its incident hyperedges) tends to belong to large hyperlinks-though not necessarily a large number of them-and thus is more likely to be infected than nodes with a low NC. When a node with a high NC becomes infected, it then tends to infect more nodes/neighbors according to the SICP spreading process. As assortativity increases by our rewirings, nodes with large NC are more likely connected by an order-3 hyperedge. However, such order-3 hyperedges could hardly further increase the infection probability of its component. Connecting nodes that already have a high chance of infection restricts the capacity of the epidemic to spread broadly across the network. This aligns with the principle that selecting multiple well-connected seed nodes limits overall prevalence \cite{abhelakh2026fairinfluencemaximizationhypergraphs}. 

\begin{figure}[h]
    \centering
    \includegraphics[width=0.49\linewidth]{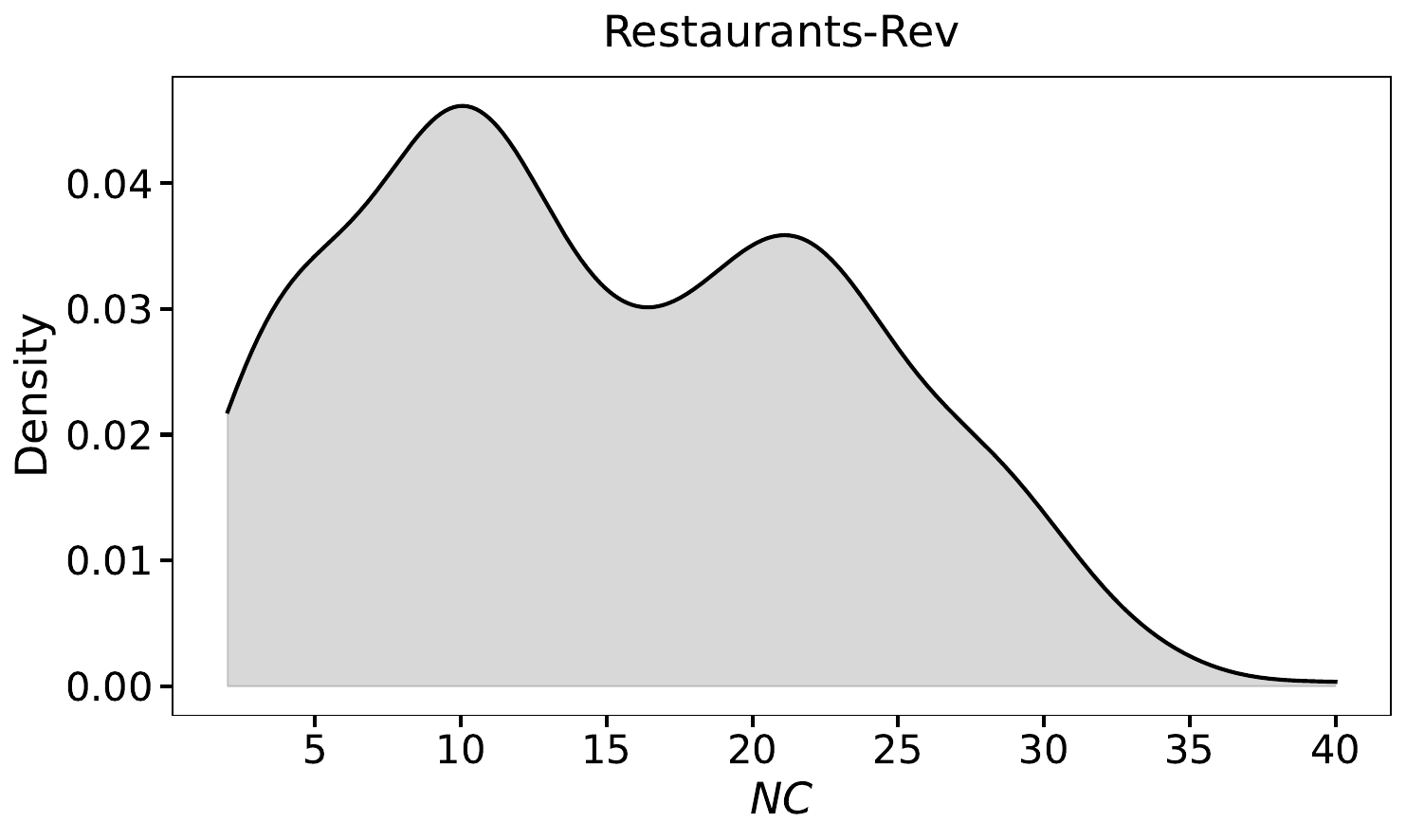}
    \includegraphics[width=0.49\linewidth]{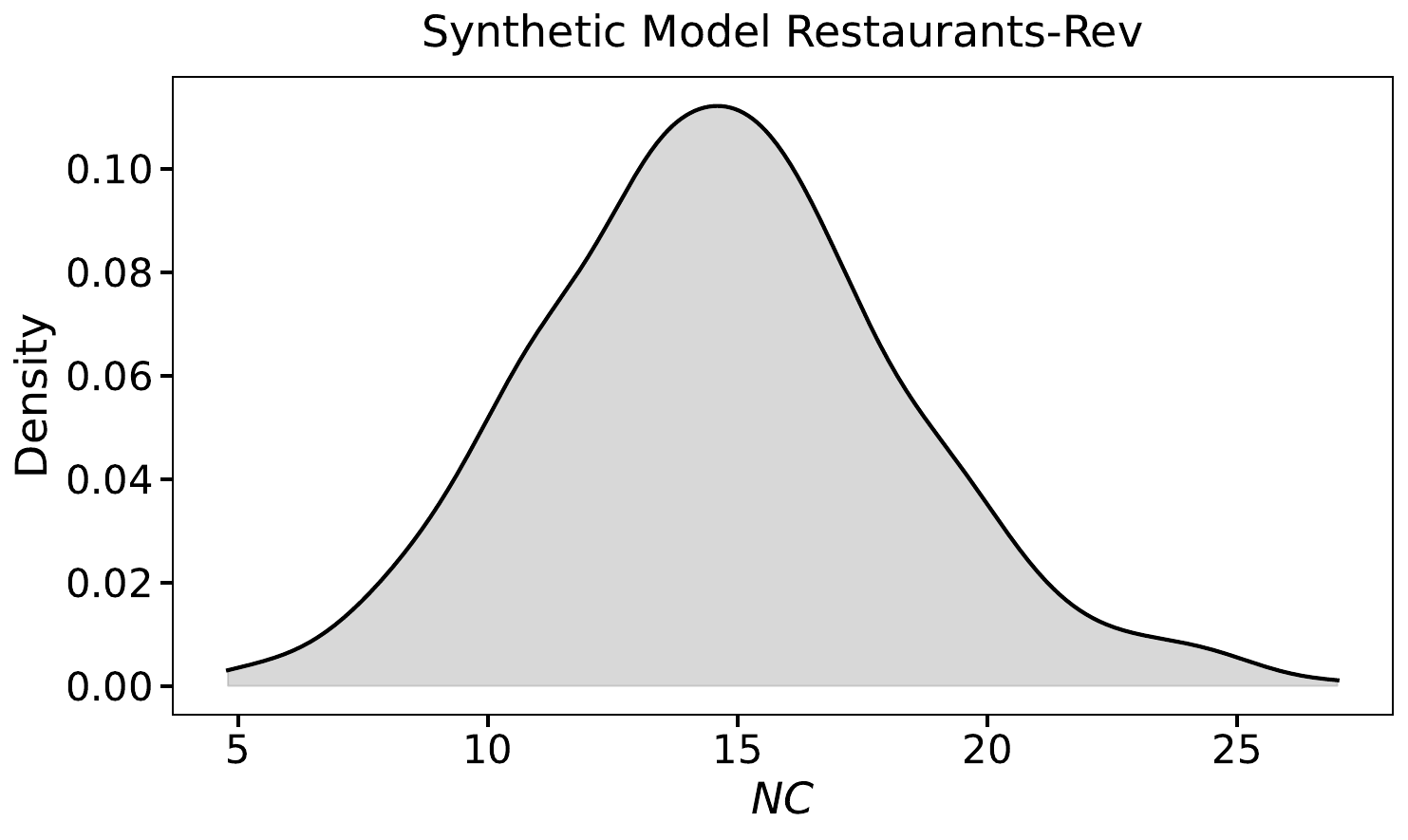}\\
    \includegraphics[width=0.49\linewidth]{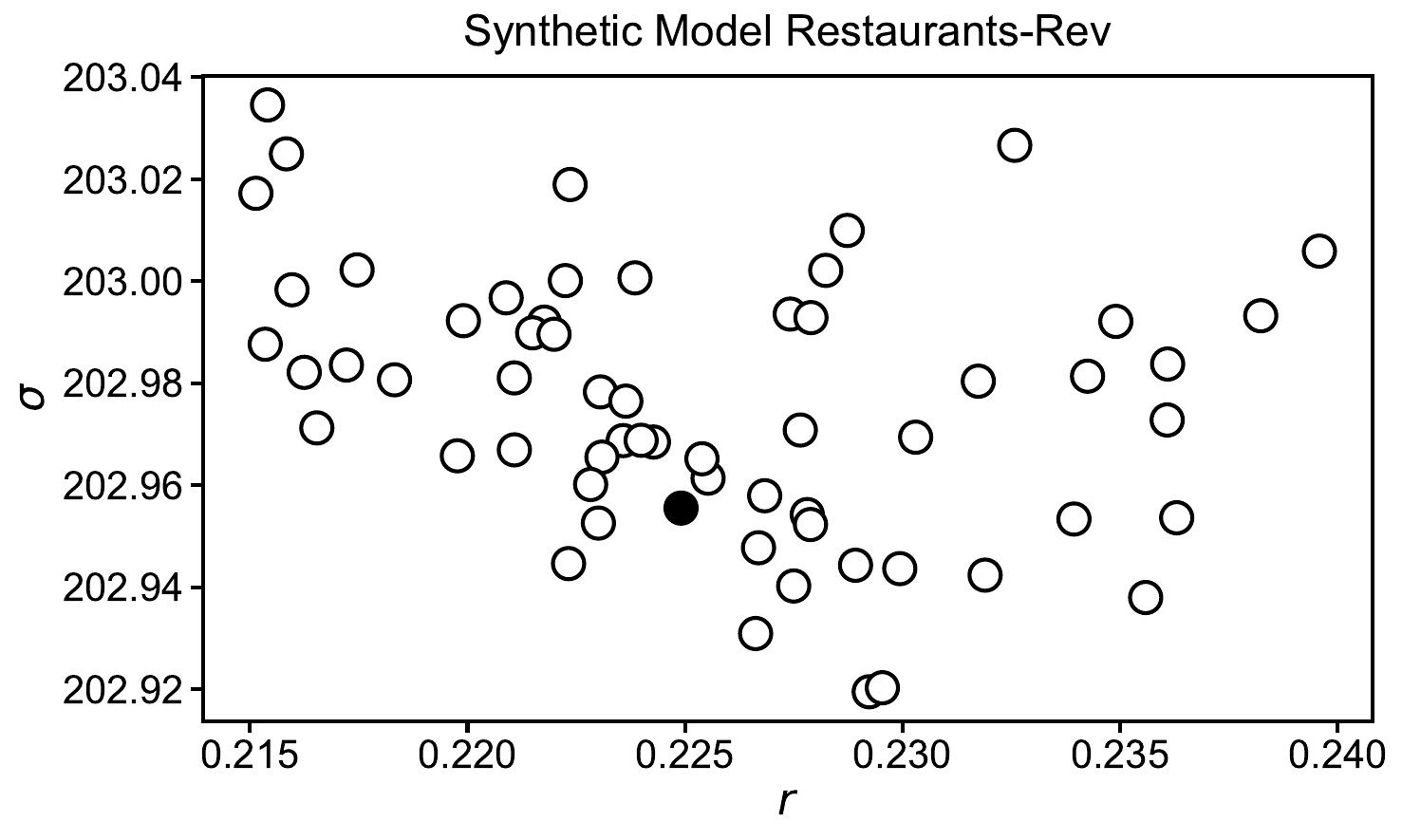}
    \includegraphics[width=0.49\linewidth]{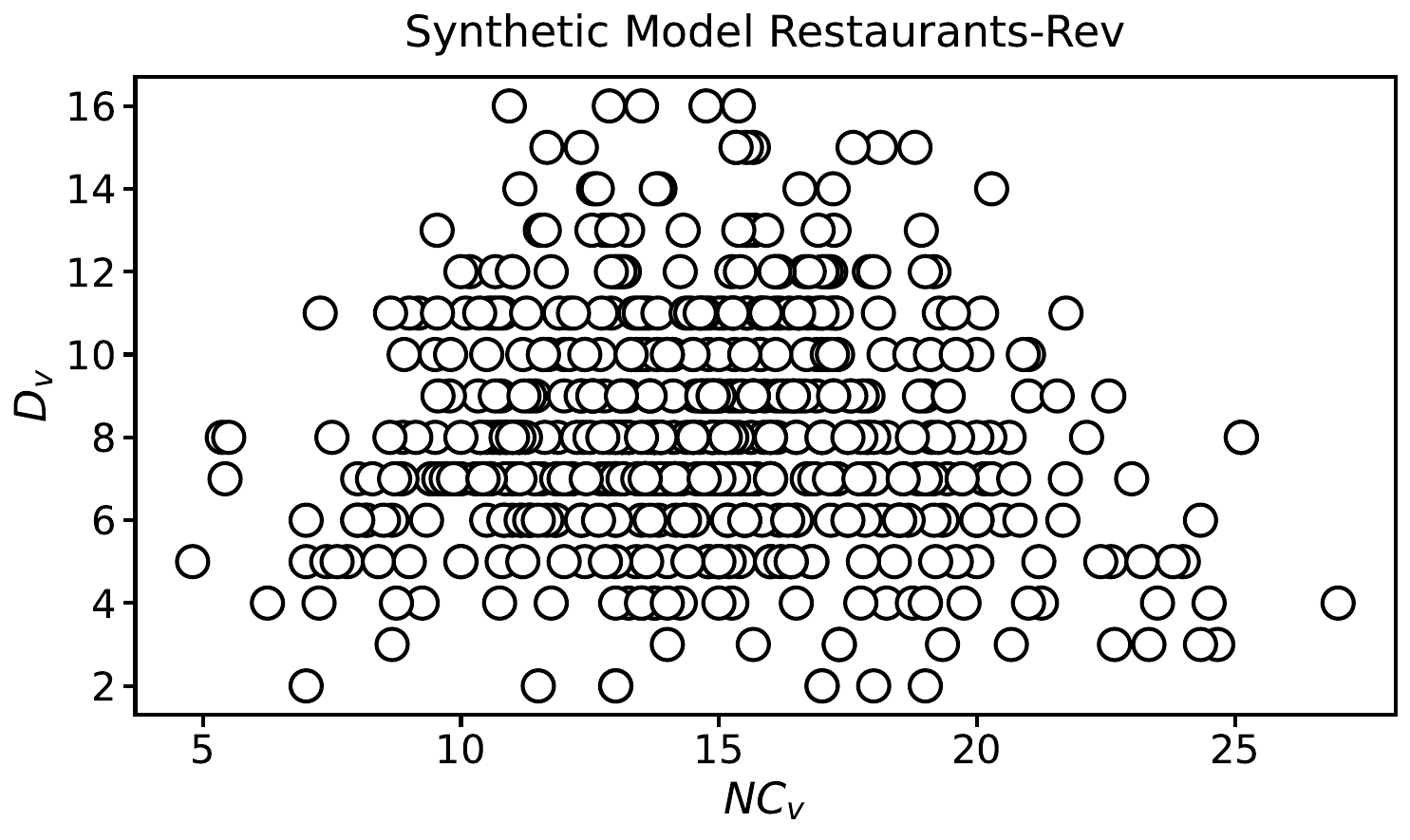}
    \captionsetup{labelfont=bf}
    \caption{\textbf{Top left:} NC distribution in the real-world network Restaurants-Rev. \textbf{Top right:} NC distribution in the synthetic Restaurants-Rev network. \textbf{Bottom left:} Prevalence $\sigma$ versus NC assortativity $r$ in the synthetic Restaurants-Rev network and its rewired counterparts. \textbf{Bottom right:} Hyperedge degree $D_v$ versus neighborhood coefficient $NC_v$ of each node in the synthetic Restaurants-Rev network.}
    \label{fig:ncdistribution_and_variation}
\end{figure}

The trend that prevalence decreases with assortativity is not observed in contact-primary-school. Given each real-world network, we have also generated a random synthetic network which preserves the number of nodes and number of hyperedges of each order and connect randomly chosen nodes to form hyperedges. This trend is not observable in most random synthetic networks either, as exemplified in Figure~\ref{fig:ncdistribution_and_variation} (bottom left panel). Hence, the influence of assortativity on prevalence is topology dependent. 

\begin{figure*}[t]
    \centering

    \includegraphics[width=0.49\textwidth]{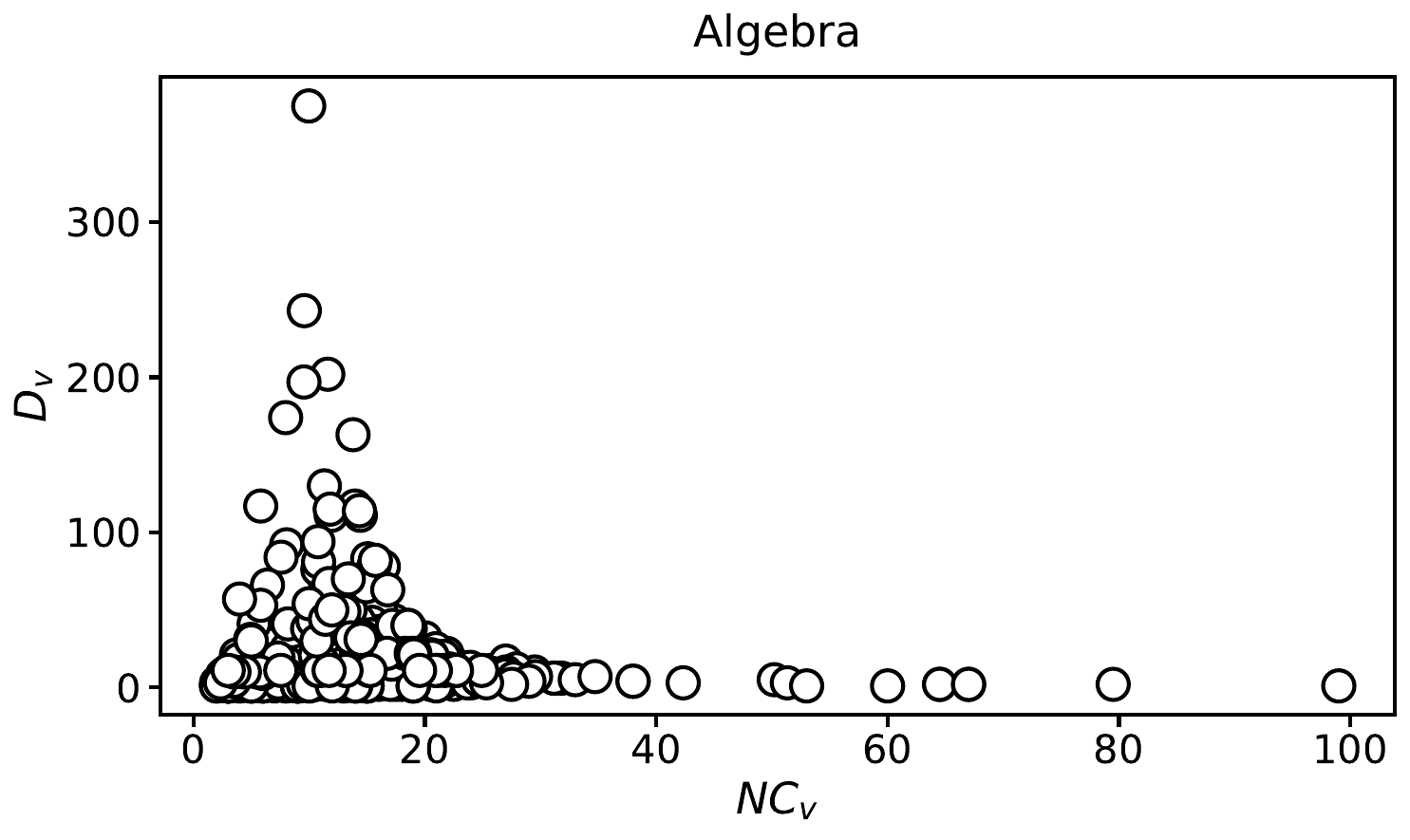}
    \includegraphics[width=0.49\textwidth]{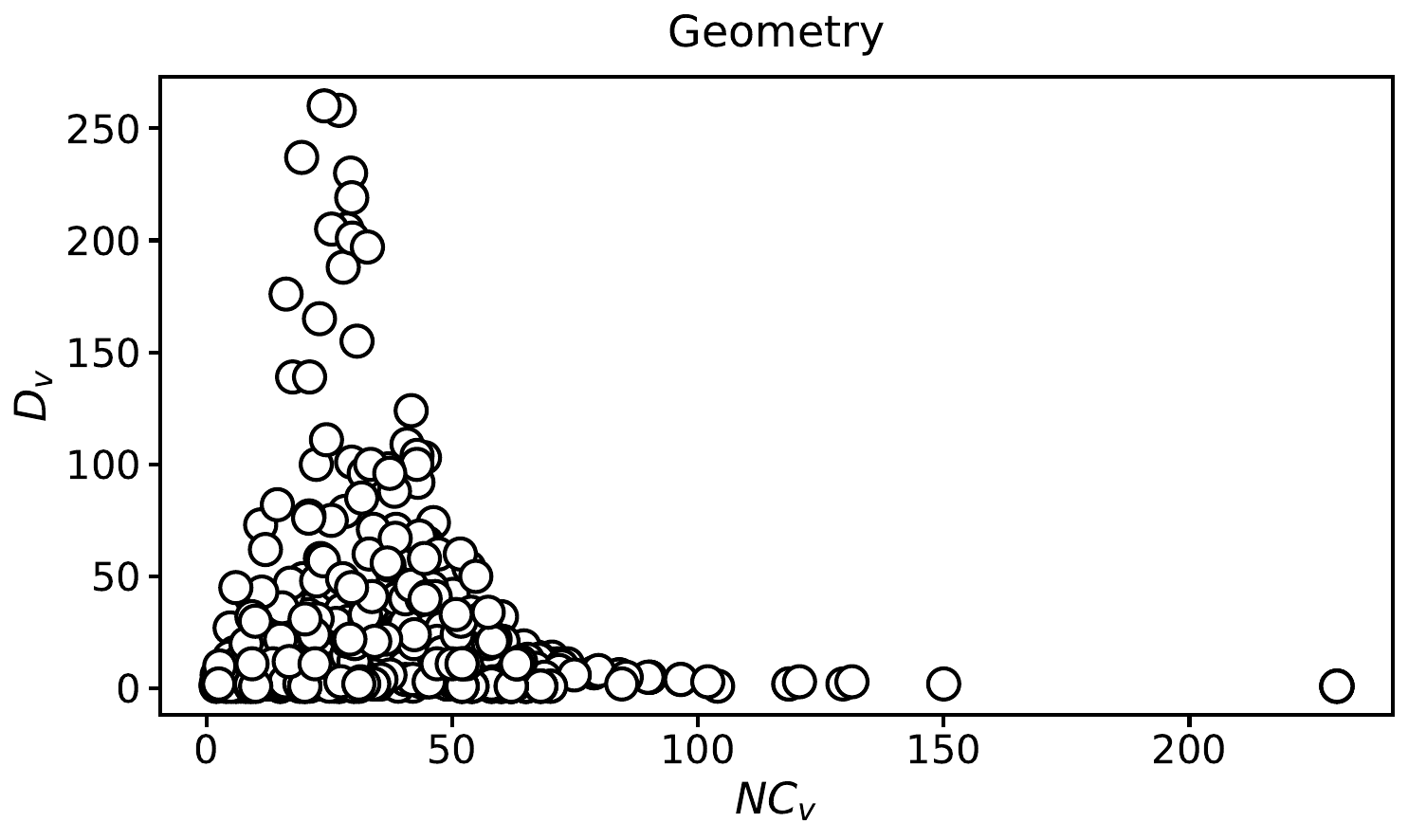}
    \vspace{0.5em}
    \includegraphics[width=0.49\textwidth]{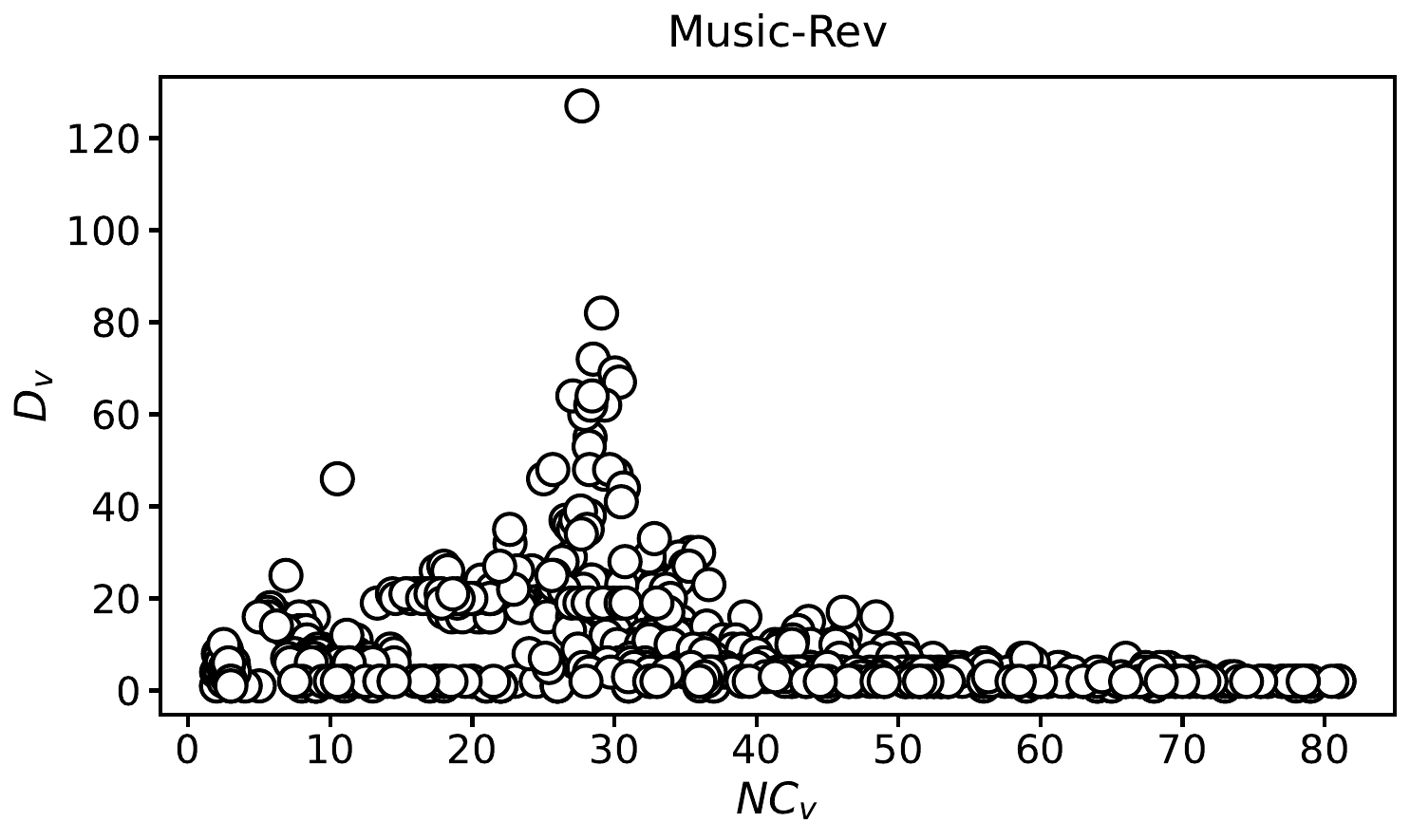}
    \includegraphics[width=0.49\textwidth]{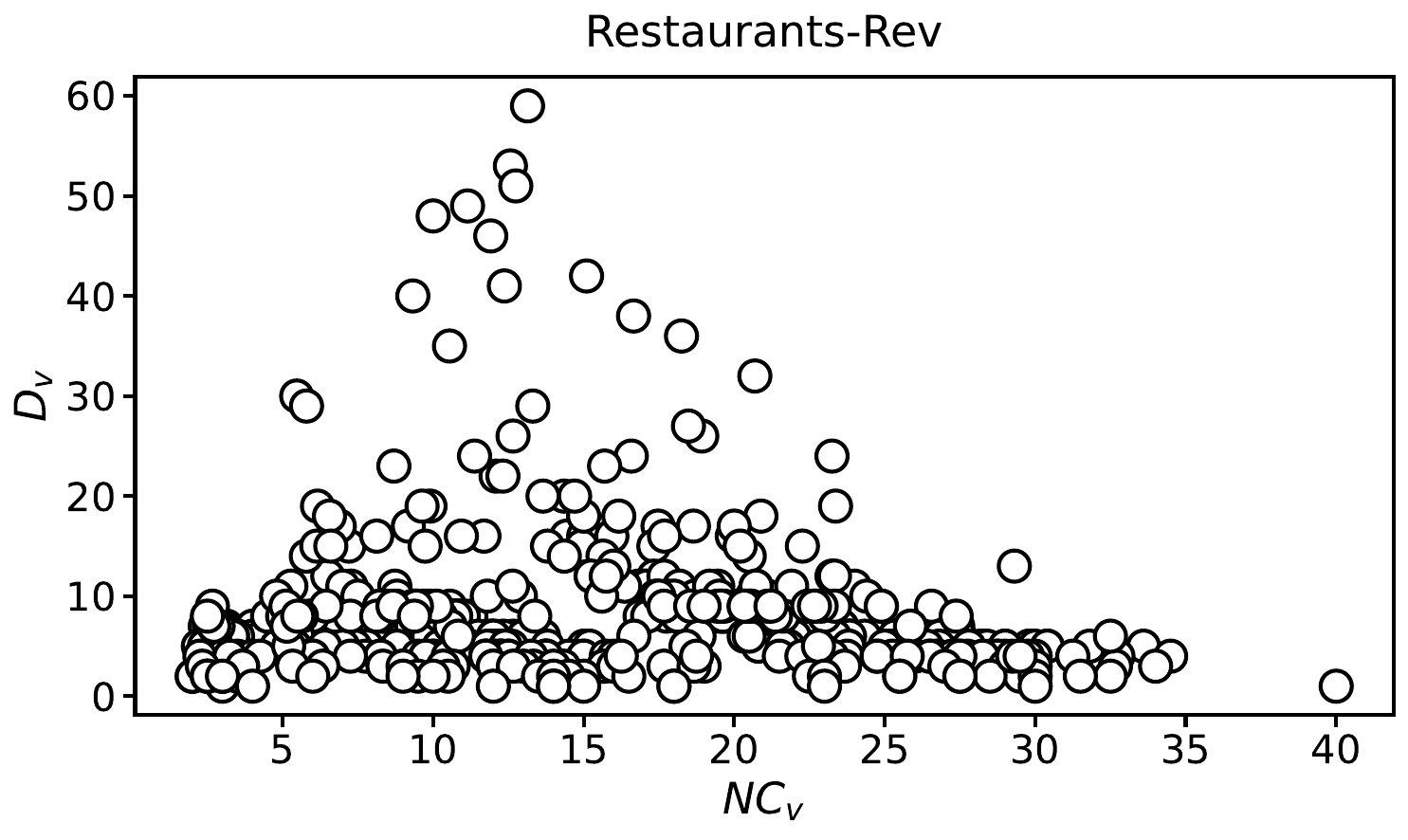}
    \vspace{0.5em}
    \includegraphics[width=0.49\textwidth]{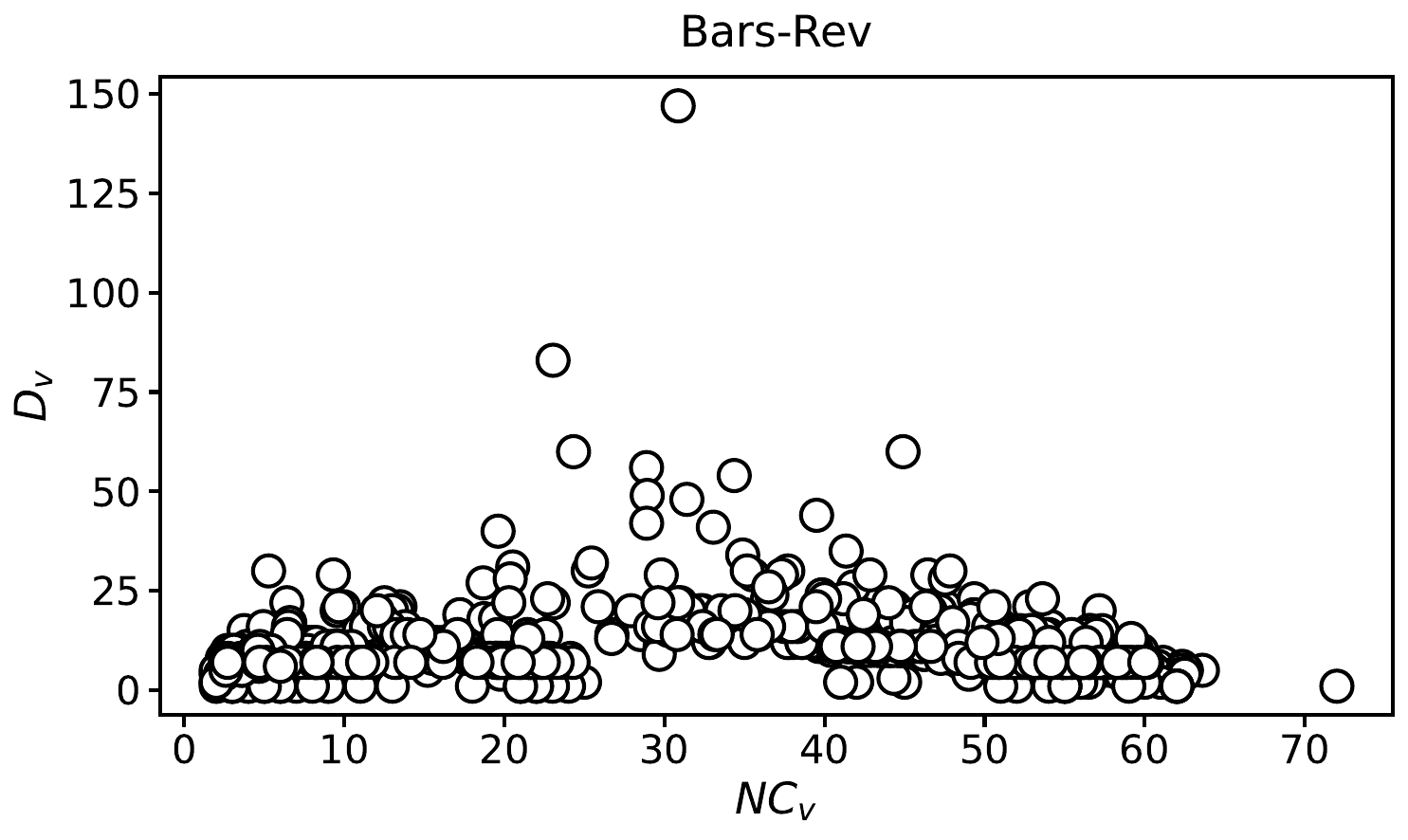}
    \includegraphics[width=0.49\textwidth]{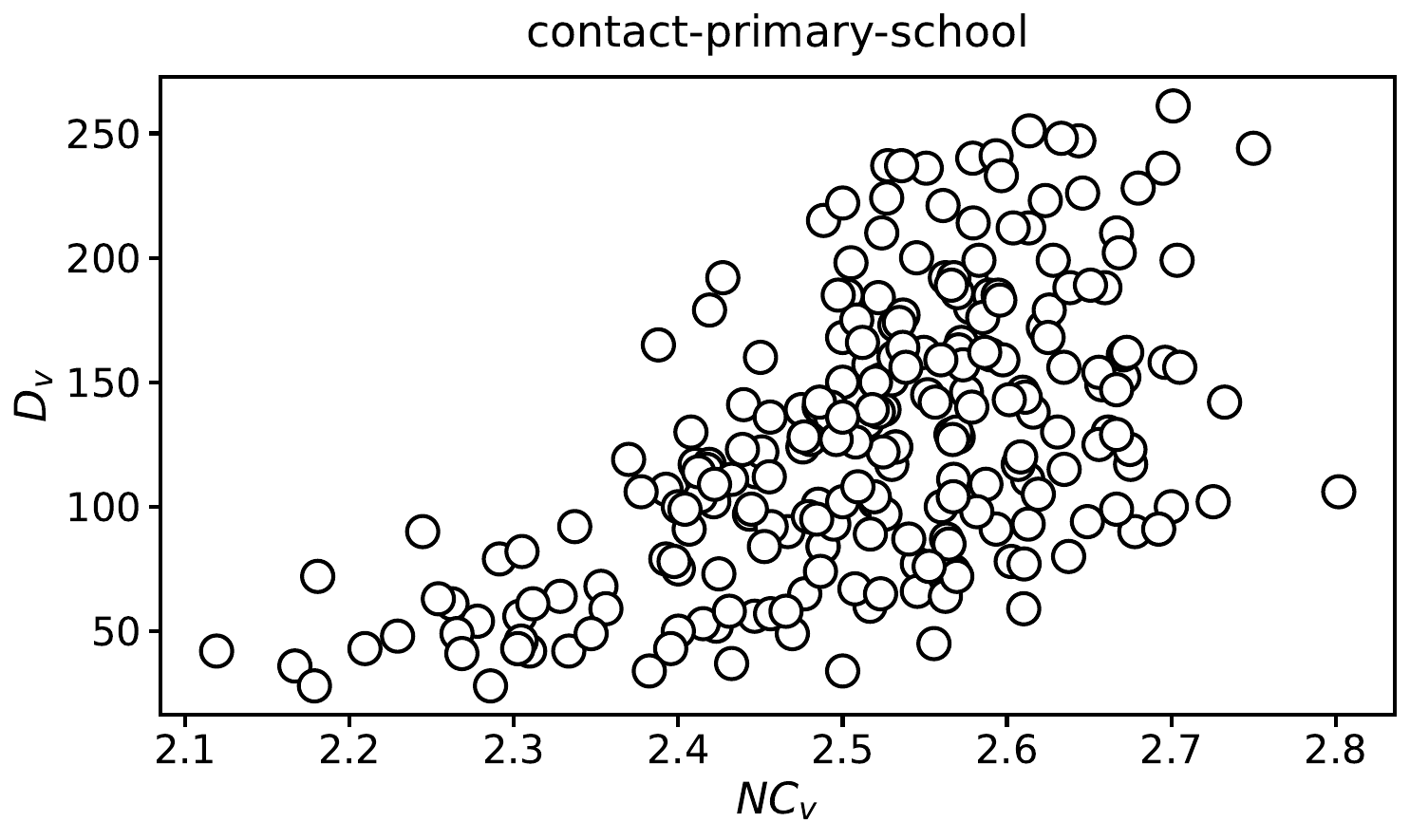}
    
    \captionsetup{labelfont=bf}
    \caption{
    Hyperedge degree $D_v$ versus neighbourhood coefficient $NC_v$ of each node $v$ in every real-world hypergraph.
    }
    \label{fig:all_datasets_nc-hdeg}
\end{figure*}

The five real-world networks in which this trend is evidently observed exhibit significant topological heterogeneity. For example, their NC distributions are non-normal, whereas the NC distributions of contact-primary-school and randomly generated hypergraphs conform to a normal distribution (see e.g., top row of Figure~\ref{fig:ncdistribution_and_variation}. Figure~\ref{fig:all_datasets_nc-hdeg} shows that these five networks feature not only a wide range of NC values, but also nodes with large NC but low hyperedge degree. Such topological heterogeneities—specifically in the NC distribution and the presence of nodes participating in few but large hyperedges—may drive the pronounced negative correlation between assortativity and prevalence. Furthermore, when a network is rewired towards either higher or lower assortativity, its structural randomness increases, which elevates the epidemic prevalence. This supports why all rewired variants of the contact-primary-school network exhibit higher prevalence than the original network, regardless of their assortativity. This also explains why the prevalence in real-world hypergraph e.g., Restaurant-Rev is lower than that in the corresponding synthetic network generated by the random model, as shown in  Figure~\ref{fig:ncdistribution_and_variation} (bottom left panel). Ultimately, determining the type of hypergraphs in which this negative influence of assortativity/rewirings on prevalence is evident, overwhelming the influence of network randomness, is an open question.

The change in prevalence values with rewiring is generally small, though the evident trend is insightful. The contribution of order-3 hyperlinks to prevalence is relatively small due to their small hyperlink size and small count. Then change in prevalence with assortativity could be more pronounce when rewirings are applied to higher-order hyperedges and when they are large in number. 

\subsection{Variation Depending on Seed node}
\begin{figure*}[t]
    \centering
    \includegraphics[width=0.49\textwidth]{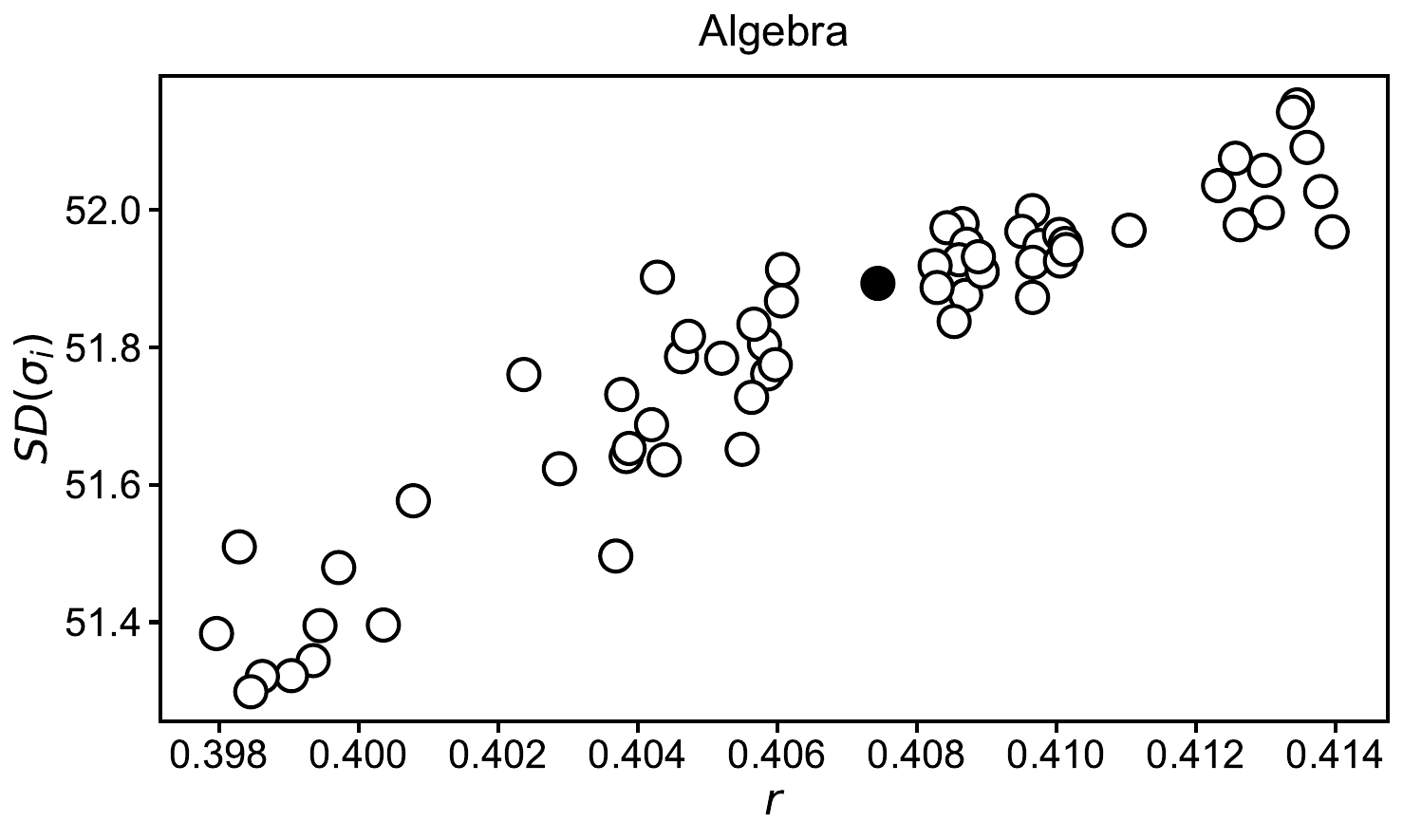}
    \includegraphics[width=0.49\textwidth]{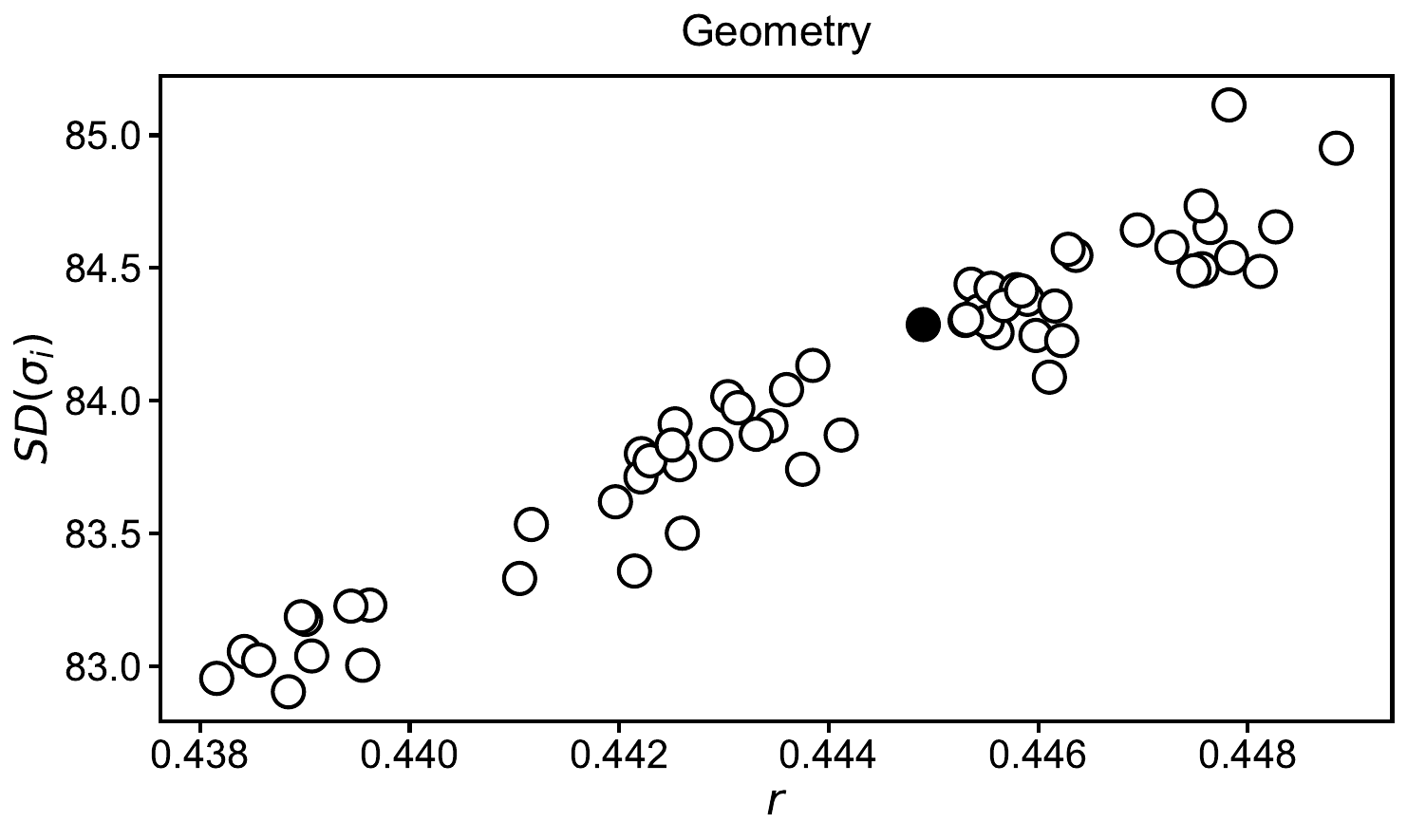}
    \vspace{0.5em}
    \includegraphics[width=0.49\textwidth]{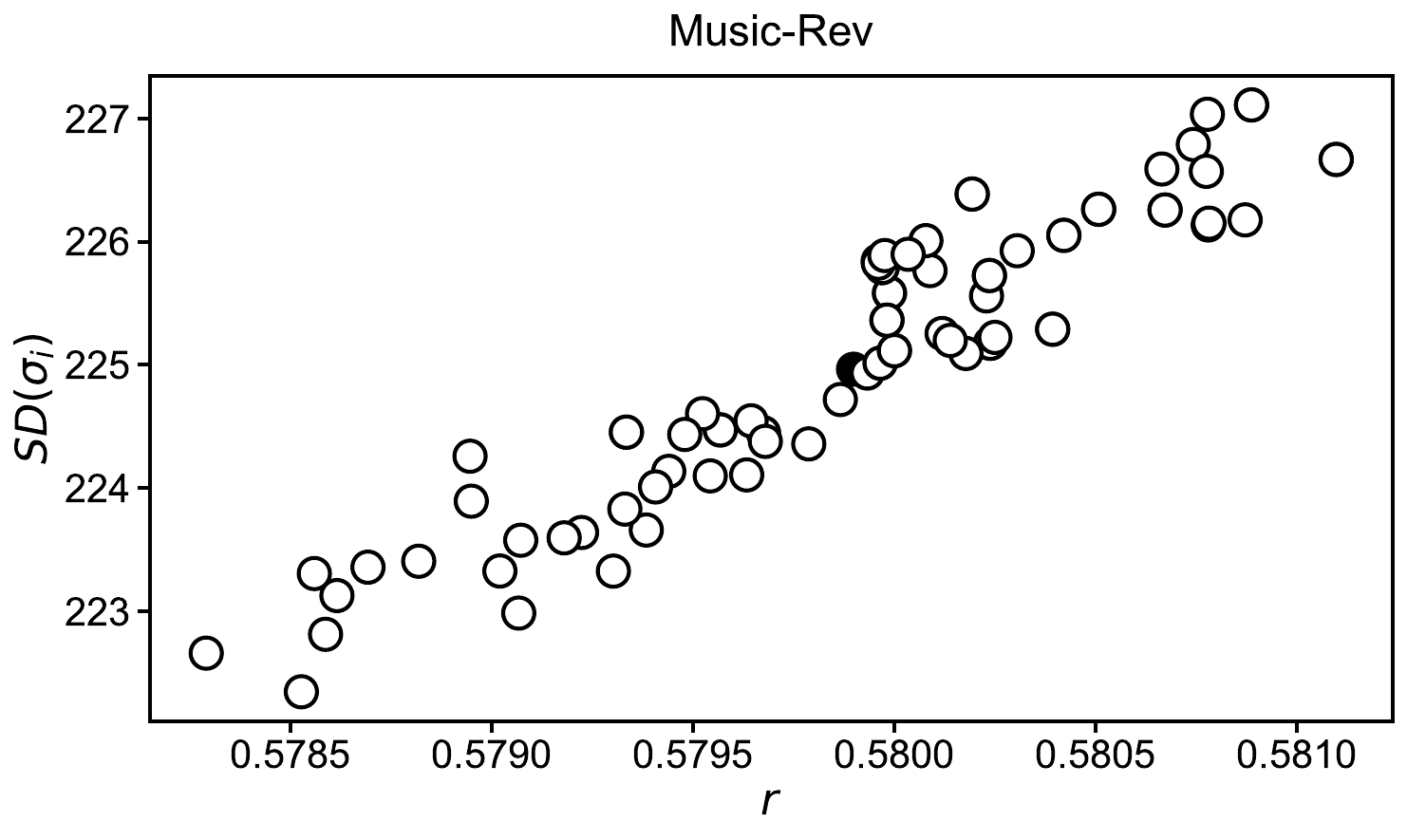}
    \includegraphics[width=0.49\textwidth]{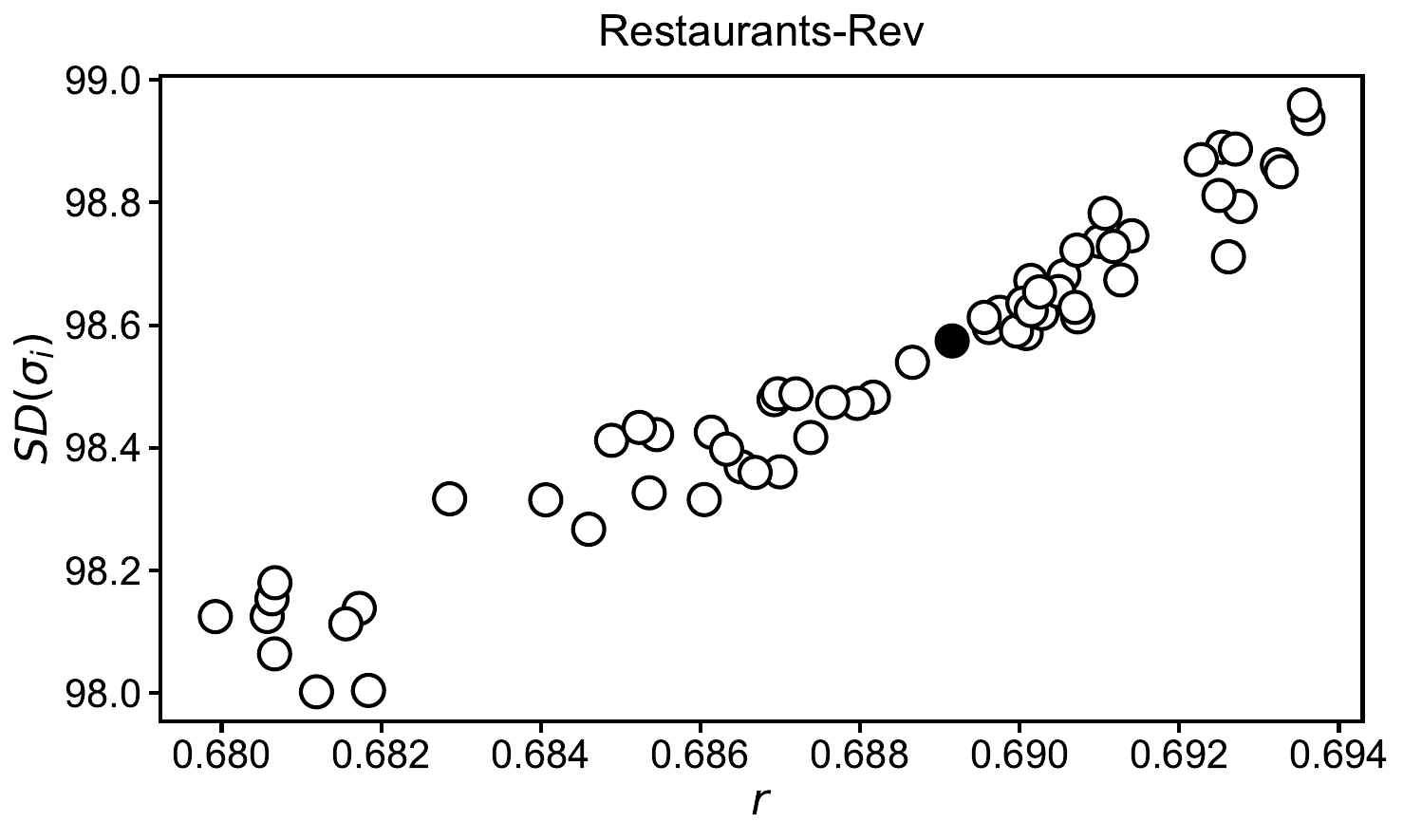}
    \vspace{0.5em}
    \includegraphics[width=0.49\textwidth]{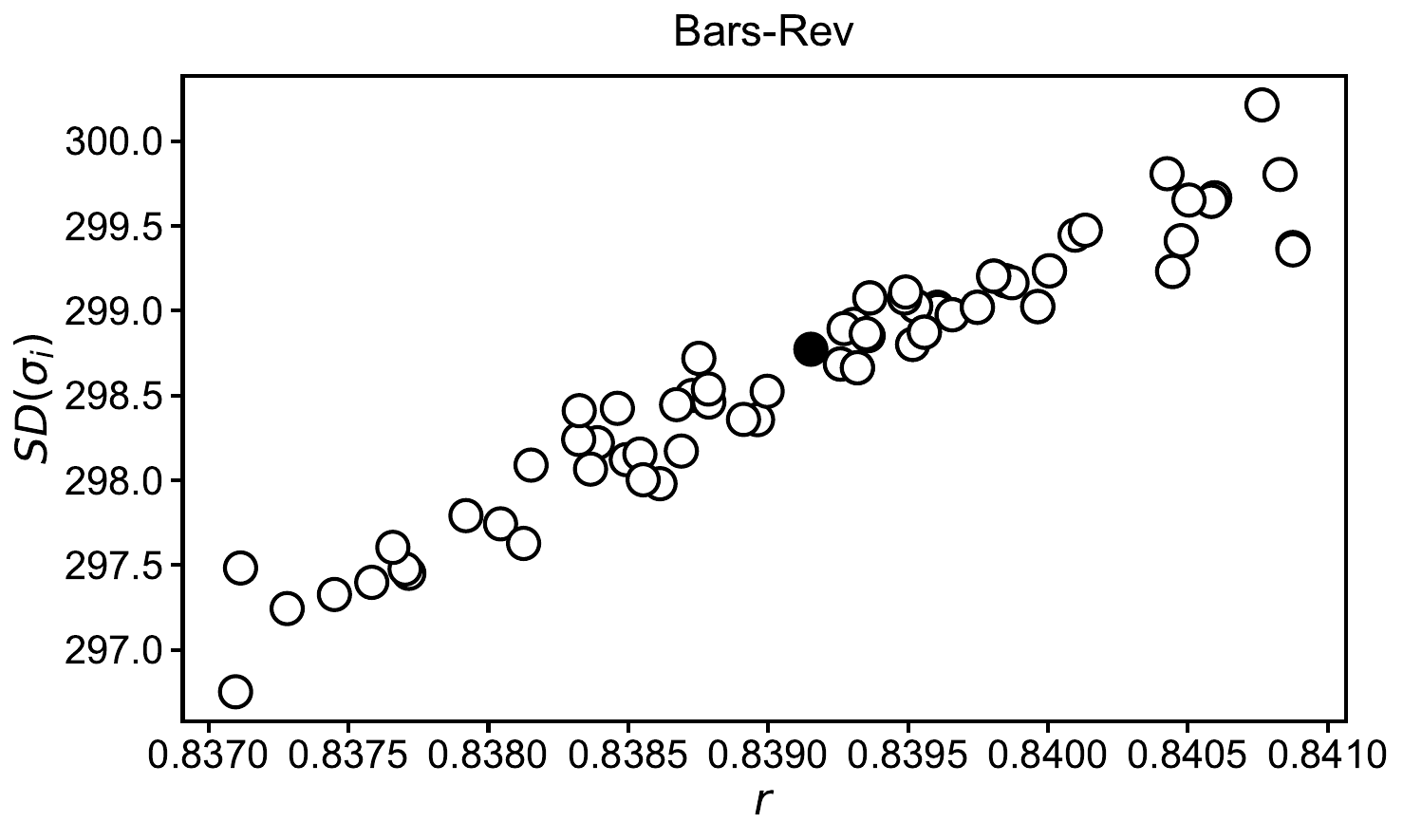}
    \includegraphics[width=0.49\textwidth]{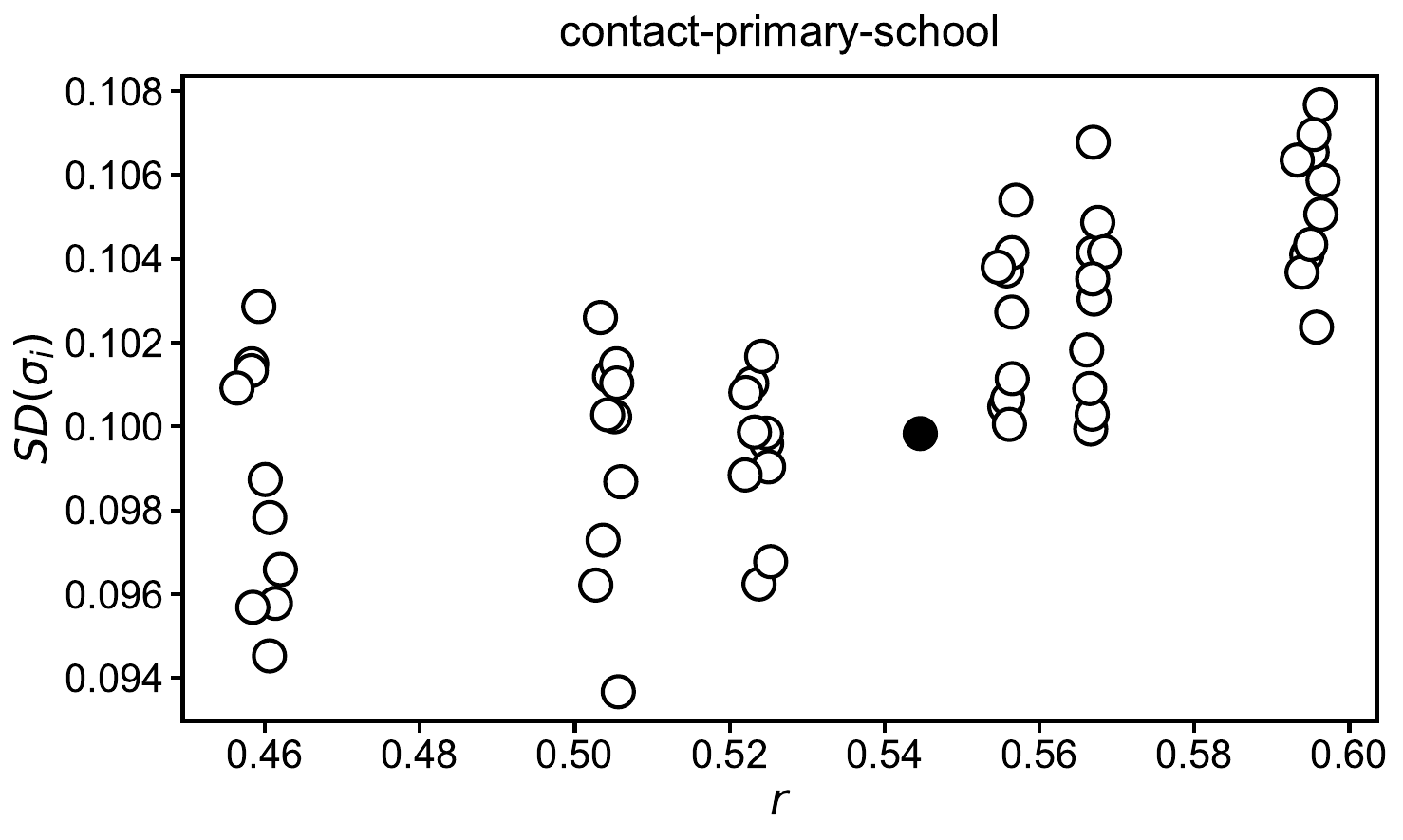}

    \captionsetup{labelfont=bf}
    \caption{
    Standard deviation of the prevalence across all seed nodes in relation to NC assortativity. The solid black dot denotes the unmodified hypergraph
    }
    \label{fig:all_datasets_stds}
\end{figure*}
So far, we have studied the prevalence $\sigma=\frac{\sum_i\sigma_i}{N}$, averaged over all possible single seed nodes. We will now study the variation in prevalence $\sigma_i$ introduced by each possible single seed node.

Figure~\ref{fig:all_datasets_stds} shows the standard deviation of $\sigma_i$ over all possible single seed nodes.
Here, we observe that the variation increases evidently with assortativity. This indicates that assortative rewiring will amplify the differences between seed nodes in prevalence. Consider a network exhibiting high assortativity. If an initial seed node has a low NC value, it infects relatively few neighbors on average, which themselves tend to possess low NC values and subsequently infect few additional nodes. Conversely, a seed node with a high NC value is capable of infecting a larger number of neighbors, which similarly tend to have high NC values and can further propagate the epidemic throughout their respective large neighborhoods.

Again, the trend in which the variance in prevalence, resulting from seed node selection, increases with assortativity is consistently observed across real-world networks—most prominently in the aforementioned five networks—whereas it remains largely absent in most random hypergraphs.

\section{Conclusions}\label{conclusion}
In this paper, we investigated how Neighborhood Coefficient (NC) assortativity of a hypergraph impacts the SICP spreading process by applying hyperdegree-preserving rewirings to real-world hypergraphs to increase or decrease their assortativity. We show that a lower NC assortativity yields higher average epidemic prevalence and lower prevalence variance across seed selections. High NC assortativity connects high-NC nodes together via hyperlinks, creating redundant infection pathways that constrain broad network spreading and amplify prevalence variance depending on whether a seed has a high or low NC. These assortativity effects are, however, mostly absent when comparing synthetic random hypergraphs to their rewired counterparts. This highlights the complexity that the influence of a topological property on dynamics processes is also topologically dependent. 

While we demonstrated our method by rewiring order-3 hyperlinks, it remains valuable to explore the rewiring of hyperlinks of other orders, whose effect on spreading process could be more pronounced if they are large in number. Furthermore, systematic studies are needed to investigate specific types of hypergraphs where these effects manifest clearly by, for example, utilizing advanced synthetic hypergraphs that control multiple topological properties simultaneously.

\section*{Acknowledgments}
We thank for the support of Netherlands Organisation for Scientific Research NWO (
project FORT-PORT no. KICH1.VE03.21.008, and NEPTARGOS no. KICH1.VE05.23.003) and NExTWORKx, a collaboration between TUDelft and KPN on future telecommunication networks.


%
%
\bibliographystyle{spmpsci}
\bibliography{refs}

\end{document}